\documentclass[prb,twocolumn,floatfix]{revtex4}

\usepackage{amssymb, amsmath}
\usepackage{mathrsfs}
\usepackage[usenames]{color}
\usepackage{subfigure, wrapfig}
\usepackage[dvips]{graphicx}

\newcommand{\deleted}[1]{{\color{Brown}$\blacksquare$}}

\newcommand{\avg}[1]{\langle #1 \rangle}
\newcommand{\op}[1]{#1}
\newcommand{\opdag}[1]{#1^\dagger}
\newcommand{\up}{\uparrow}
\newcommand{\dn}{\downarrow}

\newcommand{\Lg}{\mathcal{L}}
\newcommand{\eadb}[2]{\epsilon #1 \partial #2}
\newcommand{\eadbL}[2]{\epsilon^{\mu\nu\lambda} #1 \partial_{\nu} #2}
\newcommand{\eda}[1]{\epsilon \partial #1}
\newcommand{\edaL}[1]{\epsilon^{\mu\nu\lambda} \partial_{\nu} #1}
\newcommand{\uv}[1]{\hat{#1}}
\newcommand{\vv}[1]{\mathbf{#1}}
\newcommand{\vvsym}[1]{\pmb{#1}}
\newcommand{\Id}{\mathbb{I}}
\newcommand{\Zeros}{\mathbb{O}}
\newcommand{\Ones}{\mathbb{E}}
\newcommand{\elif}[1]{\tilde{#1}}
\newcommand{\Hopf}[3]{#1^\mu #2 \left(\frac{\epsilon_{\mu\nu\lambda} \partial^\nu}{\partial^2} \right) #3^\lambda}

\usepackage[dvips]{hyperref}

\begin{document}

\title{Doped Kagom\'{e} System as Exotic Superconductor}
\author{Wing-Ho Ko}
\author{Patrick A. Lee}
\author{Xiao-Gang Wen}
\affiliation{Department of Physics, Massachusetts Institute of Technology, Cambridge, Massachusetts 02139, USA}
\date{\today}

\begin{abstract} 
A Chern--Simons theory for the doped spin-1/2 kagom\'e system is constructed, from which it is shown that the system is an exotic superconductor that breaks time-reversal symmetry. It is also shown that the system carries minimal vortices of flux $hc/4e$ (as opposed to the usual $hc/2e$ in conventional superconductors) and contains fractional quasiparticles (including fermionic quasiparticles with \emph{semionic} mutual statistics and spin-1/2 quasiparticles with \emph{bosonic} self-statistics) in addition to the usual spin-1/2 fermionic Bougoliubov quasiparticle. Two Chern--Simons theories---one with an auxiliary gauge field kept and one with the auxiliary field and a redundant matter field directly eliminated---are presented and shown to be consistent with each other.
\end{abstract}

\maketitle 

\section{Introduction} \label{sect:intro} 

The ``perfect'' spin-1/2 kagom\'e lattice, realized recently in Herbertsmithite ZnCu$_3$(OH)$_6$Cl$_2$,\cite{Expt:Helton,Expt:Ofer,Expt:Mendels} has produced great enthusiasm in both the experimental and the theoretical condensed matter community. Experimentally, the antiferromagnetic exchange is found to be $J \approx$ 190~K, and yet no magnetic ordering is observed down to a temperature of 50~mK.\cite{Expt:Helton} Theoretically, with nearest-neighbor Heisenberg antiferromagnetic interaction, several possible ground states have been proposed, including the valance bond solid (VBS) states\cite{DSL:Hastings,VBS:Nikolic} and the Dirac spin liquid (DSL) state,\cite{DSL:Ran, DSL:Hermele} while results from exact diagonalization (ED)\cite{ED:Waldtmann} remains inconclusive as to which state is preferred.

So far both the experimental and theoretical studies have been focused on the half-filling (i.e., undoped) case. In this paper, we investigate the situation in which the kagom\'e system is doped, which could in principle be realized by substituting Cl with S. We shall take the DSL state, which at low energy is described by spin-1/2 Dirac fermions (spinons) coupled to an emergent internal gauge field, as our starting point. Naively, one might expect the system to be a Fermi liquid with small Fermi pockets opening up at the spinon Dirac nodes. However, since the system contains an emergent internal gauge field $\alpha^\mu$, filled Landau Levels (LLs) can spontaneously form. When the flux quanta of this emergent field is equal to half of the doping density, the resulting LL state is energetically favorable. (the formation of filled LLs, as induced by the internal gauge flux, has also been proposed in the case when an external magnetic field is applied to the undoped spin-1/2 kagom\'e system).\cite{DSLB:Ran} Furthermore, the strength of this $\alpha$ field and the doping density can co-fluctuate smoothly across space, resulting in a gapless excitation in density mode. Since the gapless density mode is the only gapless excitation, the LL state is actually a superconducting state. This provides an unconventional superconducting mechanism which results in a \emph{time-reversal symmetry breaking} superconductor. 

As typical for a superconductor, the state we proposed also supports electromagnetically (EM) charged vortices. In additional, since there are multiple species of emergent spinons and holons, the system also contains EM-neutral topological excitations that are analogous to quasiparticles in quantum Hall systems. To describe the superconducting state, the EM-charged vortices, and the EM-neutral quasiparticles in a unified framework, we start with the $t$--$J$ model and the DSL ansatz and construct a \emph{Chern--Simons theory}, well-known from the study of quantum Hall systems, for this system.

In our scenario, the low-energy effective theory contains four species of emergent holons, each carry a charge $e$. All four species are tied together by the emergent gauge field $\alpha^\mu$. Consequently, the flux of a minimal vortex in this superconductor is found to be $hc/4e$, as opposed to the usual $hc/2e$ in conventional superconductor. Furthermore, the quasiparticles in this scenario are shown to exhibit fractional statistics. In particular, there are fermionic quasiparticles with \emph{semionic} mutual statistics and bosonic quasiparticles carrying spin 1/2.


This paper is organized as follows: In Sect.~\ref{sect:Lg}, we derive the Chern--Simons theory starting with the $t$--$J$ model and motivate the necessity of such an ``unconventional'' formation for superconductivity. In Sect.~\ref{sect:zero-mode}, the existence of superconductivity is first explained intuitively, and then confirmed by a more rigorous derivation. The physical vortices are then discussed, with the $hc/4e$ magnetic flux explained both intuitively and mathematically. In Sect.~\ref{sect:qp_stat}, the EM-neutral quasiparticles are introduced and their statistics are derived. The discussion on these quasiparticles continue into Sect.~\ref{sect:qp_Qnum} in which their quantum numbers are analysed. In Sect.~\ref{sect:elif}, an alternative formulation of the Chern--Simons theory is presented, in which the auxiliary gauge field $\alpha^\mu$ and a redundant matter field are eliminated directly, and the results obtained are shown to be consistent with that of the previous sections. The paper concludes with Sect.~\ref{sect:conclude}.

\section{From \texorpdfstring{$t$--$J$}{t-J} Hamiltonian to Chern--Simons theory} \label{sect:Lg}

The starting point of our model for the doped kagom\'e system is the $t$--$J$ Hamiltonian:
\begin{equation} \label{eq:H_tJ}
H_{tJ} = \sum_{\avg{ij}} J \left(\vv{\op{S}}_i \cdot \vv{\op{S}}_j 
	- \frac{1}{4}\op{n}_i \op{n}_j \right)
	-t \left( \opdag{c}_{i\sigma} \op{c}_{j\sigma} + h.c. \right) \, ,
\end{equation}
where $\opdag{c}_{i\sigma}$ and $\op{c}_{j\sigma}$ are projected electron operators that forbid double occupation, and that $J > 0$. Throughout this paper we shall assume that $t > 0$ and that the system is hole doped. For $t < 0$, our results can be translated to an electron-doped system upon applying a particle-hole transformation.

Using the $U(1)$ slave-boson formulation,\cite{SB:Lee} we introduce spinon (fermion of charge 0 and spin 1/2, representing singly occupied sites) operators $\op{f}_{i\sigma}$ and holon (boson of charge $+e$ and spin 0, representing empty sites) operators $\op{h}_{i}$ such that $\opdag{c}_{i\sigma} = \opdag{f}_{i\sigma} \op{h}_i$, and apply the Hubbard--Stratonovich transformation. This yields the following partition function:
\begin{equation} \label{eq:partition_fcn}
Z = \int Df Df^\dagger Dh Dh^* D\lambda D\chi D\Delta \exp\left(-\int_0^\beta d\tau L_1\right) \, ,
\end{equation}
where
\begin{equation} \label{eq:action1}
\begin{aligned}
L_1 & = \frac{3J}{8} \sum_{\avg{ij}} (|\chi_{ij}|^2 + |\Delta_{ij}|^2)
		+ \sum_{i\sigma} f^\dagger_{i\sigma} (\partial_\tau - i \lambda_i) f_{i\sigma}\\
	& - \frac{3J}{8} \left( \sum_{\avg{ij}}\chi_{ij}^* (\sum_\sigma f^\dagger_{i\sigma}f_{j\sigma}) 
 		+ c.c. \right)\\
 	& + \frac{3J}{8} \left( \sum_{\avg{ij}}\Delta_{ij}
 		(f^\dagger_{i\up} f^\dagger_{j\dn}-f^\dagger_{i\dn} f^\dagger_{j\up}) + c.c. \right)\\
 	& + \sum_i h_i^* (\partial_\tau - i \lambda_i + \mu_B) h_i 
 		- t \sum_{\avg{ij},\sigma} h_i h_j^* f^\dagger_{i\sigma} f_{j\sigma} \, ,
\end{aligned}
\end{equation}
in which the mean-field conditions are given by $\chi_{ij} = \sum_\sigma \avg{\opdag{f}_{i\sigma} \op{f}_{j\sigma}}$ and $\Delta_{ij} = \avg{\op{f}_{i\up} \op{f}_{j\dn}-\op{f}_{i\dn} \op{f}_{j\up}}$.

Assuming mean-field ansatzes in which $\Delta_{ij} = 0$ and $\chi_{ij} = \chi e^{-i \alpha_{ij}}$, and rewriting $\lambda_i = \alpha_0^i$, we arrive at the following mean-field Hamiltonian:
\begin{equation} \label{eq:H_MF}
\begin{aligned}
H_{\textrm{MF}}\! 
& = \! \sum_{i\sigma} \opdag{f}_{i\sigma}(i \alpha_0^i - \mu_F) \op{f}_{i\sigma}
	\!-\! \frac{3\chi J}{8} \!\!\sum_{\avg{ij},\sigma}\! 
	(e^{i\alpha_{ij}}\opdag{f}_{i\sigma}\op{f}_{j\sigma} \!+\! h.c.) \\
	& + \sum_i \opdag{h}_i (i \alpha_0^i - \mu_B) \op{h}_{i}	
	- t \chi \sum_{\avg{ij}} (e^{i \alpha_{ij}}\opdag{h}_{i} \op{h}_{j} + h.c.) \, .
\end{aligned}
\end{equation}

\begin{figure}
	\subfigure[\label{fig:R_lattice}]{\includegraphics[scale=0.7]{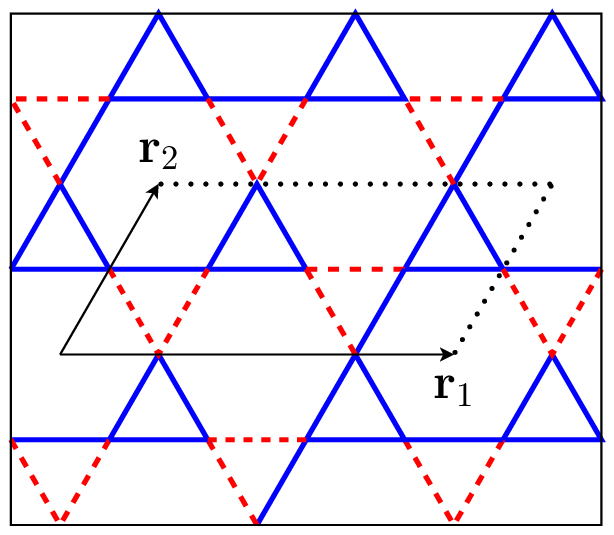}} \,
	\subfigure[\label{fig:k_lattice}]{\includegraphics[scale=0.9]{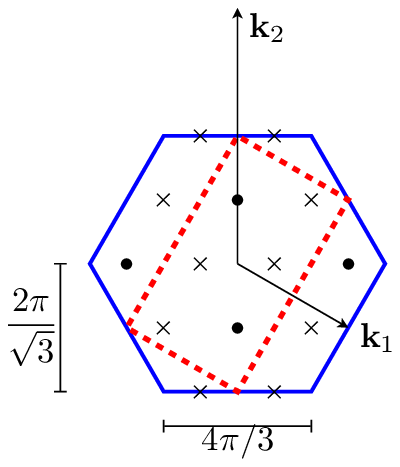}} 
	\caption{\label{fig:lattice} (a) The kagom\'e lattice with the DSL ansatz. The dashed lines correspond to bonds with $t = -1$ while unbroken lines correspond to bonds with $t = 1$. $\vv{r}_1$ and $\vv{r}_2$ are the primitive vectors of the doubled unit cell. (b) The original Brillouin zone (bounded by unbroken lines) and the reduced Brillouin zone (bounded by broken lines) of the DSL ansatz. The dots indicate locations of the Dirac nodes at half-filling while the crosses indicate locations of the quadratic minima of the lowest band. $\vv{k}_1$ and $\vv{k}_2$ are the reciprocal lattice vectors of the reduced Brillouin zone.}
\end{figure}

Observe that an internal gauge field $\alpha^\mu$ emerges naturally from this formulation. Its space components $\alpha_{ij}$ arise from the phases of $\chi_{ij}$, while its time component $\alpha_0$ arises from enforcing the occupation constraint:
\begin{equation} \label{eq:constraint}
\opdag{h}_i \op{h}_i + \opdag{f}_{i\up} \op{f}_{i\up} + \opdag{f}_{i\dn} \op{f}_{i\dn} = 1 \, .
\end{equation}

From Eq.~\ref{eq:H_MF}, it can be seen that the holons and spinons are not directly coupled with each other at the mean-field level---they are correlated only through the common gauge field $\alpha^\mu$. Consequently, if we treat $\alpha^\mu$ at the mean-field level, the spinon spectra and the holon spectra will decouple, and up to an overall energy scale both will be described by the \emph{same} tight-binding Hamiltonian.

\begin{figure}
	\subfigure[]{\includegraphics[scale=0.6]{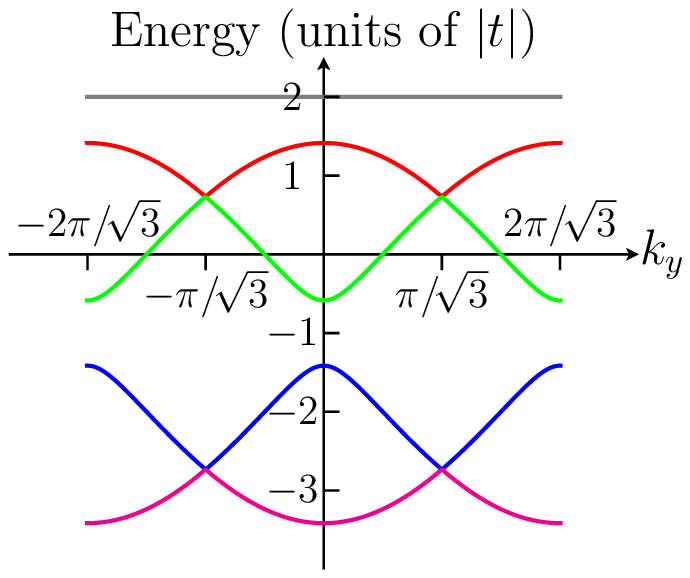}} \,
	\subfigure[]{\includegraphics[scale=0.6]{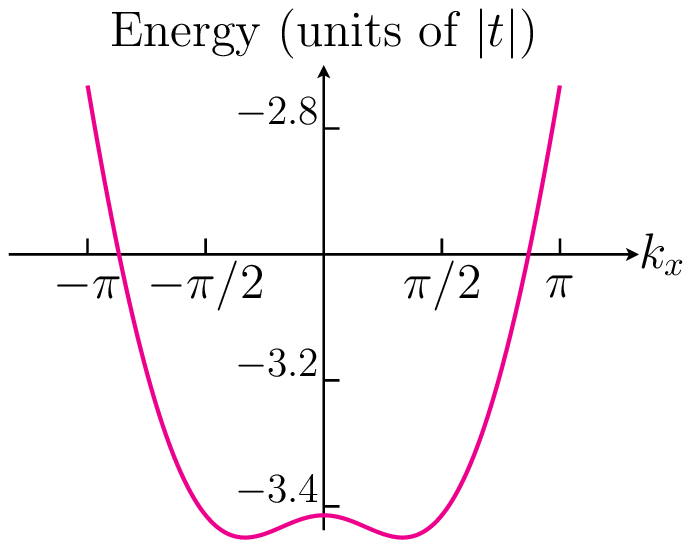}}
	\caption{\label{fig:Brillouin} The band structure of the kagom\'e lattice with the DSL ansatz (a) plotted along the line $k_x = 0$ and (b) of the lowest band plotted along the line $k_y = 0$. Note that the top band in (a) is twofold degenerate.}
\end{figure}

By gauge invariance, a mean-field ansatz for $\alpha^\mu$ is uniquely specified by the amount of fluxes through the triangles and the hexagons of the kagom\'e lattice. In particular, the DSL state is characterized by zero flux through the triangles and $\pi$ flux through the hexagons.\cite{DSL:Ran, DSL:Hermele, DSL:Hastings} By picking an appropriate gauge, the DSL state can be described by a tight-binding Hamiltonian with doubled unit cell, in which each nearest-neighbor hopping is real, has the same magnitude, but varies in sign. For the precise pattern see Fig.~\ref{fig:R_lattice}. This tight-binding Hamiltonian produces six bands, whose dispersions are, in units where the magnitude of the hopping parameter is set to $1$,
\begin{align}
E_{\textrm{top}} & = 2 \qquad \textrm{(doubly degenerate)}\\
E_{\pm,\mp} & = -1 \pm \sqrt{ 3 \mp \sqrt{2} 
	\sqrt{3-\cos 2 k_x + 2 \cos k_x \cos\sqrt{3} k_y} }
\end{align}
At any $\vv{k}$-point, $E_{-,+} \leq E_{-,-} \leq E_{+,-} \leq E_{+,+} \leq E_{\textrm{top}}$. These tight-binding bands have the following features that will be important for our purposes: (1) four degenerate shallow quadratic band bottoms in the first (lowest) band $E_{-,+}$; and (2) two degenerate Dirac nodes where the third band ($E_{+,-}$) and the fourth band ($E_{+,+}$) touches. See Fig.~\ref{fig:k_lattice} and Fig.~\ref{fig:Brillouin} for illustrations.

Now suppose the doped kagom\'e system is described by the DSL ansatz as in the undoped case, and that the doping is $x$ per site. Then each doubled unit cell will contain $6x$ holons and $3-3x$ spinons per spin. By Fermi statistics, the spinons will fill the lowest $3-3x$ bands and thus can be described by anti-spinon pockets at each Dirac node. Similarly, by Bose statistics the holons will condense at each quadratic band bottom. This state shall be referred to as the Fermi-pocket (FP) state.

However, the FP state is not the only possibility. In particular, an additional amount of uniform $\alpha$ field can be spontaneously generated to produce LLs in both the holon and spinon sector. The resulting state shall be referred to as the LL state. In the absence of holons (i.e., at half filling), both mean-field calculation and projection wavefunction study indicate that the LL state is energetically favored over the FP state.\cite{DSLB:Ran} Since the spinon bands are linear near half-filling while the lowest holon band is quadratic near its bottoms, at the mean-field level the energy gain from the spinon sector (which scales as $3/2$ power of the $\alpha$ field strength) will be larger than the energy cost in the holon sector (which scales as square of the $\alpha$ field strength) at low doping. Therefore, even after the holons are taken into account, the LL state is expected to have a lower energy than the FP state.

Furthermore, from mean-field it can be seen that the energy gain will be maximal when the $\alpha$ field is adjusted such that the zeroth spinon LLs are exactly empty. Since each flux quanta of the $\alpha$ field corresponds to one state in each LL, and that each anti-spinon pocket contains $3x/2$ states for a doping of $x$ per site, the flux must be $3x$ flux quanta per doubled unit cell for the zeroth spinon LLs to be empty. 

As for the holon sector, there are $6x$ holons per doubled unit cell or equivalently $3x/2$ holons per band bottom. Since the holon carries the electric charge and are hence are mutually repulsive, one may expect them to fill the four band bottoms symmetrically. In such case the first LL of each of the holon band bottom would be exactly \emph{half-filled}, which implies that the holons would form four Laughlin $\nu=1/2$ quantum Hall states. Since the Laughlin $\nu=1/2$ state is gapped and incompressible, this symmetric scenario should be energetically favorable.\footnote{The stability of the Laughlin $\nu=1/2$ state of boson can be seen by flux attachment argument. Since there are two flux quanta per boson, attaching one flux quanta to each boson maps the Laughlin $\nu=1/2$ state of boson to an integer quantum Hall state of fermion, which is gapped and incompressible. In contrast, a $\nu=1$ quantum Hall state for boson is mapped to a free fermion gas upon attaching one flux quanta to each boson, and hence is unstable.}

From the physical arguments given above, it can be seen that the effective description of this system is analogous to that of a (mulit-layered) quantum Hall system, and thus may contain non-trivial topological orders, manifesting in, e.g., fractional quasiparticles with non-trivial statistics. In order to describe such system, we adopt a \emph{hydrodynamic approach} well-known in the quantum Hall literature.\cite{Duality:Fisher, Duality:Wen, Duality:ZeeBook} In this approach, a duality transformation is applied, in which a gauge field is introduced to describe the current associated with a matter field, and which the two are related by:
\begin{equation} \label{eq:duality}
J^\mu = \frac{1}{2\pi} \edaL{a_\lambda} \, ,
\end{equation}
where $J^\mu$ is the current of the matter field and $a_\lambda$ is the associated gauge field. Here $\mu$, $\nu$, and $\lambda$ are spacetime indices that run from $0$ to $2$, and $\epsilon^{\mu\nu\lambda}$ is the totally antisymmetric Levi-Civita symbol. 

In this formalism, a single-layer quantum Hall system of filling fraction (a.k.a.\@ Hall number) $\nu = 1/m$ is described by the following effective Lagrangian:
\begin{equation} \label{eq:QHE_Lg}
\Lg = -\frac{m}{4\pi} \eadbL{a_\mu}{a_\lambda}
	-\frac{e}{2\pi} \eadbL{a_\mu}{A_\lambda}
	+ \ell a_\mu j_V^{\mu} 	+\ldots \,
\end{equation}
where $A^\mu$ is the external electromagnetic field and $j_V^{\mu}$ is the current density associated with particle-like excitations. The ``$\ldots$'' represents terms with higher derivatives, and hence unimportant at low energies. In particular, at the lowest order in derivatives among the terms dropped is the ``Maxwell term'': 
\begin{equation} \label{eq:QHE_Maxwell}
\Lg_{\textrm{Maxwell}} = 
-\frac{1}{2 g^2} (\partial_\mu a_\nu-\partial_\nu a_\mu)(\partial^\mu a^\nu-\partial^\nu a^\mu) \, .
\end{equation}

The effective Lagrangian Eq.~\ref{eq:QHE_Lg} can be understood by considering the equation of motion (EOM) with respect to the dual gauge field $a^\mu$. With a stationary quasiparticle at $x_0$ such that $j_V^{\mu} = (\delta(x-x_0),0,0)$, the EOM reads, in the time-component:
\begin{equation} \label{eq:QHE_EOM}
J^0 = - \frac{e\nu}{2\pi} B + \ell \nu \delta(x-x_0) + \ldots
\end{equation}
which confirms that $\nu$ indeed equals to the filling fraction $2\pi J^0/(-eB)$, and that $j_V^{\mu} = (\delta(x-x_0),0,0)$ is a source term for a quasiparticle having charge $\ell \nu$. In particular, a physical electron at $x_0$ can be associated with $j_V^{\mu} = (\delta(x-x_0),0,0)$ and $\ell = \nu^{-1}$.

Since $j_V^{\mu}$ is a source of ``charge'' in $a^\mu$, from the duality transformation Eq.~\ref{eq:duality}, it can alternatively be viewed as a source of vortex in the matter field current $J^\mu$.

The statistics of the quasiparticles can be deduced by integrating out the dual gauge field $a^\mu$ in Eq.~\ref{eq:QHE_Lg}, from which we obtained the well-known Hopf term:
\begin{equation} \label{eq:QHE_Hopf}
\Lg' = \pi \Hopf{\widetilde{j}}{\nu}{\widetilde{j}} + \ldots \, ,
\end{equation}
where $\widetilde{j}^\mu = -(e/2\pi)\epsilon^{\mu\nu\lambda} \partial^\nu A + \ell j_V^\mu $ is the sum of terms that couple linearly to $a^\mu$.

The statistical phase $\theta$ when one quasiparticle described by $\ell = \ell_1$ winds around another with $\ell = \ell_2$ can then be computed by evaluating the quantum phase $e^{i S} = e^{i\int \Lg'}$, with $\widetilde{j}^\mu = \ell_1 j_{V1}^\mu + \ell_2 j_{V2}^\mu$ being the total current produced by both quasiparticles. This yields\cite{Duality:ZeeBook} $\theta = 2 \pi \nu \,\ell_1 \ell_2$.

In particular, for the statistical phase accumulated when an electron winds around a quasiparticle of charge $\ell\nu$ to be a multiple of $2\pi$, $\ell$ must be an integer. This provides a quantization condition for the possible values of $\ell$.

For an $N$-layer quantum Hall system, Eq.~\ref{eq:QHE_Lg} generalizes to:
\begin{equation} \label{eq:mQHE_Lg}
\begin{aligned}
\Lg & = -\frac{1}{4\pi} \eadbL{a_{I\mu} K_{IJ}}{a_{J\lambda}}
	-\frac{e}{2\pi} \eadbL{q_I a_{I\mu}}{A_\lambda} \\
	& \quad + \ell_I a_{I\mu} j_V^{\mu} + \ldots \\
& = -\frac{1}{4\pi} \eadbL{\vv{a}_\mu K}{\vv{a}_\lambda} 
	-\frac{e}{2\pi} \eadbL{(\vv{q} \cdot \vv{a}_\mu)}{A_\lambda} \\
	& \quad + (\vvsym{\ell} \cdot \vv{a}_\mu) j_V^{\mu} + \ldots
\end{aligned}
\end{equation}
here $a_I^\mu$ is the dual gauge field corresponding to the matter field in the $I$-th layer, $\vv{a}_\mu = (a_1^\mu, \ldots, a_N^\mu)^T$ and $\vv{q}= (q_1, \ldots, q_N)^T $ are $N$-by-$1$ vectors, $\vvsym{\ell}=(\ell_1, \ldots, \ell_N)^T$ is an $N$-by-$1$ \emph{integer} vector, and $K = [K_{IJ}]$ is an $N$-by-$N$ real symmetric matrix. On the second line of Eq.~\ref{eq:mQHE_Lg} and henceforth, we adopt a condensed notation in which the boldface and dot-product always refer to the vector structure in the ``layer'' indices and never in the spacetime indices.

In the multi-layer case, assuming that $\det{K} \neq 0$, the procedure for integrating out the dual gauge fields can similarly be carried out, which yields:
\begin{equation} \label{eq:mQHE_Hopf}
\Lg' = \pi \Hopf{(\widetilde{\vv{j}}^T)}{K^{-1}}{\widetilde{\vv{j}}} + \ldots \, .
\end{equation}
where $\widetilde{\vv{j}}^\mu = - \vv{q} (e/2\pi)\epsilon^{\mu\nu\lambda} \partial^\nu A  + \vvsym{\ell} j_V^\mu$. The statistical phase $\theta$ when one quasiparticle described by $\vvsym{\ell} = \vvsym{\ell}_1$ winds around another with $\vvsym{\ell} = \vvsym{\ell}_2$ can then be computed in a similar way as in the single-layer case, which yields $\theta = 2\pi \; \vvsym{\ell}_1^T K^{-1} \vvsym{\ell}_2$. The information of quasiparticle statistics is thus contained entirely in $K^{-1}$.

Except for the complication that there is both an external EM field $A^\mu$ and an internal constraint gauge field $\alpha^\mu$, the doped kagom\'e system we proposed is completely analogous to a multi-layer quantum Hall system. We shall therefore construct a Chern--Simons theory similar to that of Eq.~\ref{eq:mQHE_Hopf} by assigning a dual gauge field to each species of matter field. 

For the holon sector, we can represent the holons at each of the four band bottoms by a dual gauge field $b_J^\mu$ $(J = 1,2,3,4)$. Since the holons at each band bottom form a Laughlin $\nu = 1/2$ state, the total Hall number for the holon sector is $\sum_J \nu_J = 2$. For the spinon sector the situation is more subtle. Since the zeroth LL is empty and all the LLs below it are fully filled at each Dirac node, we may represent the spinons near each of the four Dirac nodes by a dual gauge field $a_I^\mu$ $(I=1,2,3,4)$ having Hall number $\nu = -1$. However, since $\alpha$ is internal the combined system of holons and spinons must be $\alpha$ neutral, which requires $\sum_{\textrm{all species}} \nu = 0$ and hence in the spinon sector $\sum_I \nu_I = -2$. To circumvent this problem, we introduce two additional dual gauge fields $a_5^\mu$ and $a_6^\mu$, each having Hall number $\nu = +1$. The two fields $a_5^\mu$ and $a_6^\mu$ can be thought of as arising from the physics of spinons near the band bottoms of the two spin species. In this setting, $a_1^\mu, \ldots, a_4^\mu$ are expected to carry good spin and $\vv{k}$ quantum numbers,\footnote{The $\vv{k}$ quantum numbers should be regarded as center-of-mass crystal momentum of the Hall condensate.} while $a_5^\mu$ and $a_6^\mu$ are expected to carry good spin quantum number only. Note also that $a_1^\mu, \ldots, a_4^\mu$ possess an emergent $SU(4)$ symmetry of spin and pseudo-spin (i.e., $\vv{k}$-points).  

Assembling the different species, the low-energy effective theory for the doped kagom\'e system is given by the following Chern--Simons theory:
\begin{widetext}
\begin{align}
\label{eq:Lg_full}
\Lg & = \frac{1}{4\pi} \sum_{I=1}^4 \eadbL{a_{I\mu}}{a_{I\lambda}}
	- \frac{1}{4\pi} \sum_{I=5}^6 \eadbL{a_{I\mu}}{a_{I\lambda}}
	- \frac{2}{4\pi} \sum_J	\eadbL{b_{J\mu}}{b_{J\lambda}}
	+ \frac{1}{2\pi} \eadbL{\left(\sum_I a_{I\mu} + \sum_J b_{J\mu}\right)}
		{\alpha_{\lambda}} \notag \\
	& \qquad + \frac{e}{2\pi} \sum_J \eadbL{b_{I\mu}}{A_{\lambda}}
	+ \left(\sum_I \ell_I a_{I\mu} + \sum_J \ell_J b_{J\mu}\right) j_V^\mu 
	+ \ldots \\ \label{eq:Lg}
	& = - \frac{1}{4\pi} \eadbL{\vv{c}_\mu^T K}{\vv{c}_\lambda}
	+ \frac{e}{2\pi} \eadbL{(\vv{q} \cdot \vv{c}_\mu)}{A_\lambda}
	+ (\vvsym{\ell} \cdot \vv{c}_\mu) j_V^\mu + \ldots \, .
\end{align}
As before, the ``$\ldots$'' denotes terms higher in derivatives, including first and foremost the Maxwell term analogous to Eq.~\ref{eq:QHE_Maxwell}. In the second line, we have combined the eleven gauge fields internal to the system into a column vector $\vv{c}^\mu = (\alpha^\mu; a_1^\mu, \ldots, a_6^\mu; b_1^\mu, \ldots, b_4^\mu)^T$. Note that unlike Eq.~\ref{eq:mQHE_Lg}, we have included the internal gauge field $\alpha^\mu$ in $\vv{c}^\mu$. This is because $\alpha^\mu$ is internal and can be spontaneously generated while the EM field in the usual quantum Hall case is external and fixed. This distinction is crucial, as will be evident soon. The ``charge vector'' $\vv{q}$ in this case is $\vv{q} = (0;0,0,0,0,0,0;1,1,1,1)^T$, and the $K$-matrix $K$ takes the block form: 
\begin{equation} \label{eq:Kmatrix}
K = \left( \begin{array}{c|cccc|cc|cccc}
 	0 & -1 & -1 & -1 & -1 & -1 & -1 & -1 & -1 & -1 & -1 \\ \hline
 -1 & -1 &  0 &  0 &  0 &  0 &  0 &  0 &  0 &  0 &  0 \\
 -1 &  0 & -1 &  0 &  0 &  0 &  0 &  0 &  0 &  0 &  0 \\
 -1 &  0 &  0 & -1 &  0 &  0 &  0 &  0 &  0 &  0 &  0 \\
 -1 &  0 &  0 &  0 & -1 &  0 &  0 &  0 &  0 &  0 &  0 \\ \hline
 -1 &  0 &  0 &  0 &  0 &  1 &  0 &  0 &  0 &  0 &  0 \\
 -1 &  0 &  0 &  0 &  0 &  0 &  1 &  0 &  0 &  0 &  0 \\ \hline
 -1 &  0 &  0 &  0 &  0 &  0 &  0 &  2 &  0 &  0 &  0 \\
 -1 &  0 &  0 &  0 &  0 &  0 &  0 &  0 &  2 &  0 &  0 \\
 -1 &  0 &  0 &  0 &  0 &  0 &  0 &  0 &  0 &  2 &  0 \\
 -1 &  0 &  0 &  0 &  0 &  0 &  0 &  0 &  0 &  0 &  2
\end{array} \right) \, .
\end{equation}
\end{widetext}

The three terms in Eq.~\ref{eq:Lg} can be understood as follows: the first term describes smooth internal dynamics of the system; the second term describes its response under an external EM field; and the third term describes the topological excitations of the system, which can be thought of as combinations of vortices in various matter-field components. As in Eq.~\ref{eq:mQHE_Lg}, $\vvsym{\ell}$ must be an integer vector. Furthermore, since the $\alpha$ field is not a dual gauge field and contains no topological excitation (otherwise the local constraint Eq.~\ref{eq:constraint} will be violated), the $\alpha$-component of $\vvsym{\ell}$ for a physical topological excitation must be zero. 

As in the original quantum Hall case, The coefficients that appear in $K$ and $\vv{q}$ can be understood by considering the EOMs resulting from it. Upon variations with respect to $a_I^\mu$, $b_J^\mu$, and $\alpha^\mu$, we get:
\begin{align}
J_{aI}^\mu & = - \frac{1}{2\pi} \edaL{\alpha_\lambda} 
	\qquad \textrm{(I = 1,2,3,4)}\, , \label{eq:spinon_EOM}\\
J_{aI}^\mu & = \frac{1}{2\pi} \edaL{\alpha_\lambda} 
	\phantom{-}\qquad \textrm{(I = 5,6)}\, , \label{eq:spinon_EOM2}\\
J_{bJ}^\mu & = \frac{1}{2} \cdot \frac{1}{2\pi} \edaL{\alpha_\lambda} + 
	\frac{e}{2\pi} \edaL{A_\lambda} \, , \label{eq:holon_EOM}\\
0 & = \sum_I J_{aI}^\mu + \sum_J J_{bJ}^\mu \, . \label{eq:alpha_EOM}
\end{align}
The first three equations are in agreement with the picture that spinons form integer quantum Hall states while holons form Laughlin $\nu = 1/2$ states under the presence of $\alpha$ flux, and that spinons carry no EM charge while holons carry EM charge $e$. Moreover, the fourth equation can be seen as a restatement of the occupation constraint Eq.~\ref{eq:constraint}.

For brevity, we shall introduce two abbreviations henceforth. First, we shall omit spacetime indices that are internally contracted. Hence we shall write $\eadb{a}{b}$ instead of $\eadbL{a_\mu}{b_\lambda}$ and $(\eda{a})^\mu$ instead of $\edaL{a_\lambda}$. In a similar spirit, we shall write $\partial a \partial a$ instead of $(\partial_\mu a_\nu - \partial_\nu a_\mu)(\partial_\mu a_\nu - \partial_\nu a_\mu)$ for the Maxwell term. Second, we shall write vectors and matrices in block form whenever appropriate, which we abbreviate by using $\Id_n$ to denote an $n$-by-$n$ identity matrix, $\Zeros_{m,n}$ to denote an $m$-by-$n$ zero matrix, and $\Ones_{m,n}$ to denote an $m$-by-$n$ matrix with all entries equal to $1$ (such that $c\Ones_{m,n}$ denotes an $m$-by-$n$ matrix with all entries equal to $c$). In this notation, the $\vv{q}$-vector becomes $\vv{q} = (0;\Zeros_{1,4},\Zeros_{1,2},\Ones_{1,4})^T$ and the $K$-matrix in Eq.~\ref{eq:Kmatrix} becomes:
\begin{equation}
K = \begin{pmatrix} \label{Kmatrix_brief}
  0 & -\Ones_{1,4} 	& -\Ones_{1,2} 	& -\Ones_{1,4} \\
-\Ones_{4,1} 	&  -\Id_4& \Zeros_{4,2}	& \Zeros_{4,4} \\
-\Ones_{2,1} 	& \Zeros_{2,4} 	&   \Id_2& \Zeros_{2,4} \\
-\Ones_{4,1} 	& \Zeros_{4,4} 	& \Zeros_{4,2}	&   2\Id_4
\end{pmatrix} \, .
\end{equation}

\section{Superconducting Mode and physical vortices} \label{sect:zero-mode}

\begin{figure}
\includegraphics[scale=0.75]{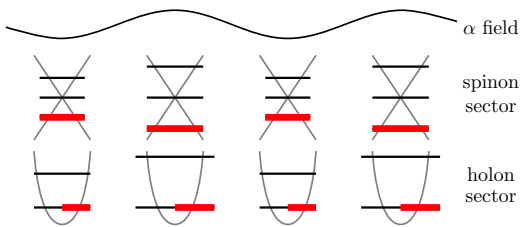}
	\caption{\label{fig:breathing} The physical picture of the breathing mode. The filled LL states are indicated by thick (red) horizontal lines while the unfilled LL states are indicated by the thin (black) horizontal lines. The original band structure for spinon and holon when no additional $\alpha$ flux is also indicated in the background (gray).}
\end{figure}

Usually, the formation of LLs will imply that all excitations are gapped. However, this is true only if the gauge field is external (i.e., fixed). \emph{Since the $\alpha$ field is internal, smooth density fluctuations can occur while keeping the local constraint Eq.~\ref{eq:constraint} and the LL structure intact}. Intuitively, if the $\alpha$ field varies across space at a sufficiently long wavelength, then the spinons and holons in each local spatial region can still be described by the LL picture, but the LLs will have a larger (smaller) spacing in regions where the $\alpha$ field is stronger (weaker). Since the LL structure is intact and the wavelength of this variation can be made arbitrarily long, the energy cost of such ``breathing mode'' can be made arbitrarily small. This breathing mode is thus a gapless charge-density mode of the system. See Fig.~\ref{fig:breathing} for illustration. Note that all species of holons and spinons co-fluctuate with the $\alpha$ field in this density mode. A similar binding mechanism in the context of cuprates is proposed in Ref.~\onlinecite{Pairing:Lee}. 

The other excitations of the system can be grouped into two general types. The first type consists of smooth density fluctuations in which the fluctuations of holons, spinons, and $\alpha$ field are mismatched. The second type consists of quasiparticle excitations that involve holons or spinons excited from one LL to another. Both types of excitations are gapped.  \emph{Since the breathing mode is the only gapless mode, it is non-dissipative, and hence the system is a superfluid when the coupling to EM fields are absent. Moreover, since the breathing mode includes the fluctuations of holons, it is charged under the EM field. Hence, the system will be a superconductor when the coupling to EM field are included}.\footnote{The gapless mode described here can be considered as the Goldstone mode associated with a spontaneous symmetry broken ground state (c.f.\@ Ref.~\onlinecite{DSLB:Ran}). With this association, the superconductivity can be seen as arising from the usual Anderson--Higgs mechanism in which this Goldstone mode is ``eaten up'' by the electromagnetic field.} Note that this superconductor breaks the time-reversal symmetry, since the sign of the additional amount of $\alpha$ flux is flipped under time reversal. Furthermore, since all four species of holons are binded together in the breathing mode, each carrying charge $+e$, a minimal vortex in this superconductor is expected to carry a flux of $hc/4e$. We shall now show these claims more vigorously from the Chern--Simons theory Lagrangian we derived in Eq.~\ref{eq:Lg}.

It is easy to check that the $K$-matrix $K$ in Eq.~\ref{eq:Kmatrix} contains \emph{exactly one zero eigenvalue}, with eigenvector $\vv{p}_0 = (2;-2\Ones_{1,4}, 2\Ones_{1,2}; \Ones_{1,4})^T$. Let $\lambda_i$ be the eigenvalues of $K$, with $\vv{p}_i$ the corresponding eigenvectors, let $P = [\vv{p}_0,\vv{p}_1,\ldots,\vv{p}_{10}]$ be the orthogonal matrix form by the eigenvectors of $K$, and let $\vv{c}' = (c'_0, \ldots, c'_{10})^T = P^\dagger \vv{c}$. Then, Eq.~\ref{eq:Lg} can be rewritten in terms of $\vv{c}'$ as:
\begin{equation} \label{eq:Lg'}
\begin{aligned}
\Lg & = -\frac{1}{4\pi} \sum_{j>0} \lambda_j \eadb{c'_j}{c'_j} 
	+ \frac{e}{2\pi} \eadb{(\vv{q} \cdot P \vv{c}')}{A}
	+ (\vvsym{\ell} \cdot P \vv{c}')_\mu j_V^\mu \\
	& \quad + g \partial c'_0 \partial c'_0 + \ldots \\
	& = \frac{e}{2\pi} (\vv{q} \cdot \vv{p}_0) \eadb{c'_0}{A}
	+ (\vvsym{\ell} \cdot \vv{p}_0) c'_{0\mu} j_V^\mu + g \partial c'_0 \partial c'_0 + \ldots \\ 
	& \quad + \textrm{(terms without $c'_0$)} \, .
\end{aligned}\end{equation}
The Maxwell term $g \partial c'_0 \partial c'_0$ for $c'_0$ in Eq.~\ref{eq:Lg'} originates from the terms in ``$\ldots$'' of Eq.~\ref{eq:Lg}, which is ordinarily suppressed by the Chern--Simons terms. However, since the Chern--Simons term $\eadb{c}{c}$ vanishes for $c'_0$, the \emph{Maxwell terms term becomes the dominant term} for $c'_0$ at low-energy and in the absence of external EM fields. Note that although the $\alpha$ field itself does not have a Maxwell term (since it arises from an occupation constraint), the zero-mode $c'_0$ does have a Maxwell term originated from the matter-field components.

Since the Maxwell term has a gapless spectrum, we see that the zero-mode $c'_0$ indeed corresponds to a gapless excitation. Moreover, since all other gauge-field components have non-zero Chern--Simons terms, excitations in these gauge-field components are gapped (these excitations corresponds to the ``mismatched'' density fluctuation mentioned earlier), verifying the earlier assertion that there is only one gapless density mode. Moreover, since $\vv{q}\cdot \vv{p}_0 \neq 0$, we see that the zero-mode is indeed charged under the external EM field. Hence, as argued above, the doped system is a superconductor.\footnote{The existence of the zero-mode (and hence superconductivity) is in fact a rather general consequence of zero total Hall number (i.e., $\sum_{\textrm{all species}} \nu = 0$). See Ref. \onlinecite{Duality:Wen}.}

The eigenvector $\vv{p}_i$ can be interpreted as the ratio of density fluctuations between the different field components in the mode $c'_i$. Thus the zero-mode indeed involves the fluctuations of all species of spinons and holons, tied together by the internal $\alpha$ field.

Since the system is a superconductor, when a sufficiently large external B field is applied, physical vortices, with the amount of flux through each vortex quantized, are expected to form. In the Chern--Simons formulation, these physical vortices manifest in the topological term (i.e., the $(\vvsym{\ell} \cdot \vv{c})_\mu j_V^\mu$ term) in Eq.~\ref{eq:Lg}. Taking an isolated topological excitation with $(j_V^0, j_V^1, j_V^2) = (\delta(x-x_0), 0, 0)$, considering the EOM associated with $c'_0$ as resulted from Eq.~\ref{eq:Lg'}, and remembering that $(\eda{A})^0 = \epsilon^{0\mu\nu} \partial_\mu A_\nu = B$ is the physical magnetic field, we obtain (in units which $\hbar = c = 1$): 
\begin{equation} 
\label{eq:Bflux} 
B = -\frac{2\pi}{e} \frac{\vvsym{\ell} \cdot \vv{p}_0} 
{\vv{q} \cdot \vv{p}_0} \delta(x-x_0) + \ldots \, .
\end{equation} 
This is the \emph{Meissner effect}, which again confirms that the system is a superconductor. Moreover, it is easy to check that non-zero $|(\vvsym{\ell}\cdot\vv{p}_0)/(\vv{q}\cdot\vv{p}_0)|$ has a minimum of $1/4$ (attained by, e.g., an $\vvsym{\ell}$-vector having a single ``$+1$'' in one of its $b_J$ components and ``$0$'' in all its other components). From this we conclude that the magnetic flux through a minimal vortex is $hc/4e$, justifying the intuitive claim given above.

\section{Quasiparticles---Statistics} \label{sect:qp_stat}

It is important to note that not all topological excitations are EM-charged. The structure of these EM-neutral topological excitations highlights the differences between this system and a conventional superconductor, and hence qualify the adjective ``exotic.'' We shall call these EM-neutral topological excitations ``quasiparticles,'' to distinguish them from the EM-charged ``physical vortices'' considered in the previous section.

From Eq.~\ref{eq:Bflux}, a topological excitation carries a non-zero magnetic flux if and only if $\vvsym{\ell} \cdot \vv{p}_0 \neq 0$. In other words, a topological excitation is EM-neutral if and only if it does not couple to the zero-mode. Note that quantity $\vvsym{\ell} \cdot \vv{p}_0$ can be regarded as the zero-mode ``charge'' carried by the topological excitation. A topological excitation with $\vvsym{\ell} \cdot \vv{p}_0 \neq 0$ couples to the zero-mode and carries its ``charge,'' which induces an $1/r$ ``electric'' field of the zero-mode and gives rise to a diverging energy gap $\Delta \sim \ln L$, where $L$ is the system size. In comparison, a topological excitation that satisfies $\vvsym{\ell} \cdot \vv{p}_0=0$ is decoupled from the zero-mode and hence has a finite energy gap and short ranged interactions. These EM-neutral topological excitations are thus analogous to the (possibly fractionalized) quasiparticles in quantum Hall systems, and it is sensible to consider the (mutual) statistics between them.

Recall that the set of $\vvsym{\ell}$-vectors (which may have non-zero $\alpha$-component) form an eleven dimensional vector space. The set of $\vvsym{\ell}$-vectors satisfying $\vvsym{\ell} \cdot \vv{p}_0 = 0$  forms a ten dimensional subspace of this eleven dimensional space. The $K$-matrix restricted to this subspace, $K_r$, is invertible. Hence we can integrate out the gauge fields associated with this subspace (i.e., the gauge fields $c'_1,\ldots,c'_{10}$ in Eq.~\ref{eq:Lg'}). This will convert the terms we omitted in Eq.~\ref{eq:Lg'} under the texts ``terms without $c'_0$'' into a Hopf term. Explicitly, upon integrating out $c'_1,\ldots,c'_{10}$ the Lagrangian takes the form:
\begin{equation} \label{eq:Hopf} \begin{aligned}
\Lg'' & = \frac{e}{2\pi} (\vv{q} \cdot \vv{p}_0) \eadb{c'_0}{A}
	+ (\vvsym{\ell} \cdot \vv{p}_0) c'_{0\mu} j_V^\mu 
	+ g \partial c'_0 \partial c'_0 + \ldots \\
& \quad + \pi \Hopf{(\widetilde{\vv{j}}^T)}{K_r^{-1}}{\widetilde{\vv{j}}} 
	+ \ldots \\
& = (\textrm{terms with $c'_0$}) + 
	\pi \Hopf{(\widetilde{\vv{j}}^T)}{K_r^{-1}}{\widetilde{\vv{j}}}
	 + \ldots \, ,
\end{aligned} \end{equation}
(c.f.\@ Eq.~\ref{eq:mQHE_Hopf}), where $\widetilde{\vv{j}}^\mu = j_V^\mu \vvsym{\ell} +(e/2\pi) (\eda{A})^\mu \vv{q}$.

As in the quantum Hall case, from Eq.~\ref{eq:Hopf} the statistical phase $\theta$ when one quasiparticle described by $j_V^\mu \vvsym{\ell}$ winds around another described by ${j'_V}^\mu \vvsym{\ell}'$ can be read off as $\theta = 2 \pi \,\vvsym{\ell}^T K_r^{-1} \vvsym{\ell}'$. For identical quasiparticles, $\theta/2$ gives the statistical phase when two such quasiparticles are exchanged.

For explicit computation a basis for $\vvsym{\ell}$-vectors for this ten-dimensional subspace must be specified. Naively one may simply choose this basis to be the set of eigenvectors of $K$ having non-zero eigenvalues. This choice turns out to be inconvenient as some of the eigenvectors of $K$ are non-integer while the quantization condition requires all $\vvsym{\ell}$ to be integer vectors. Hence, instead we shall use the following basis:

\begin{equation}
\label{eq:lbasis}
\begin{aligned}
\vvsym{\ell}_{1} & = (0; -1,1,0,0,\Zeros_{1,2};\Zeros_{1,4})^T \, , \\
\vvsym{\ell}_{2} & = (0; -1,0,1,0,\Zeros_{1,2};\Zeros_{1,4})^T \, ,\\
\vvsym{\ell}_{3} & = (0; -1,0,0,1,\Zeros_{1,2};\Zeros_{1,4})^T \, , \\
\vvsym{\ell}_{4} & = (0; \Zeros_{1,4},\Zeros_{1,2};0,0,1,-1)^T \, , \\
\vvsym{\ell}_{5} & = (0; \Zeros_{1,4},\Zeros_{1,2};0,1,0,-1)^T \, , \\
\vvsym{\ell}_{6} & = (0; \Zeros_{1,4},\Zeros_{1,2};1,0,0,-1)^T \, , \\
\vvsym{\ell}_{7} & = (0; 0,1,0,0,\Zeros_{1,2};0,1,1,0)^T \, , \\
\vvsym{\ell}_{8} & = (0; 1,0,0,0,1,0;\Zeros_{1,4})^T \, , \\
\vvsym{\ell}_{9} & = (0; \Ones_{1,4},\Ones_{1,2}; \Ones_{1,4})^T \, , \\
\vvsym{\ell}_{10} & = (-1; \Zeros_{1,4},0,1;\Zeros_{1,4})^T \, .
\end{aligned}
\end{equation}

It can be shown that all integer $\vvsym{\ell}$-vectors satisfying $\vvsym{\ell}\cdot\vv{p}_0 = 0$ can be written as \emph{integer} combinations of the above basis vectors. It should be remarked that $\vvsym{\ell}_{1}$ through $\vvsym{\ell}_{6}$ are indeed eigenvectors of $K$, with $\vvsym{\ell}_{1}$ through $\vvsym{\ell}_{3}$ having eigenvalue $-1$ and $\vvsym{\ell}_{4}$ through $\vvsym{\ell}_{6}$ having eigenvalue $2$. However, $\vvsym{\ell}_{7}$ through $\vvsym{\ell}_{10}$ are not eigenvectors of $K$. 

In this basis, $K_r^{-1}$ takes the form:
\begin{equation} \label{eq:invKr}
K_r^{-1} \!=\! \left( \begin{array}{ccccccc|ccc} 
-2 & -1 & -1 & & & & -1 & 1 & & \\
-1 & -2 & -1 & \multicolumn{3}{c}{\raisebox{0ex}[0cm][0cm]{$\Zeros_{3,3}$}} & 0 & 1 
	& \multicolumn{2}{c}{\Zeros_{3,2}}\\
-1 & -1 & -2 & & & & 0 & 1 & &  \\
& & & 1 & 1/2 & 1/2 & 1/2 & 0 & & \\
\multicolumn{3}{c}{\raisebox{0ex}[0cm][0cm]{$\Zeros_{3,3}$}} & 1/2 & 1 & 1/2 & 1/2 & 0 
	& \multicolumn{2}{c}{\Zeros_{3,2}}\\
& & & 1/2 & 1/2 & 1 & 0 & 0 & & \\
-1 & 0 & 0 & 1/2 & 1/2 & 0 & 0 & 0 & 0 & 0 \\ \hline
1 & 1 & 1 & 0 & 0 & 0 & 0 & 0 & 0 & 0 \\
& & & & & & 0 & 0 & 0 & 1\\
\multicolumn{3}{c}{\raisebox{1.5ex}[0cm][0cm]{$\Zeros_{2,3}$}}
	& \multicolumn{3}{c}{\raisebox{1.5ex}[0cm][0cm]{$\Zeros_{2,3}$}}
	& 0 & 0 & 1 & 1
\end{array} \right).
\end{equation}

Note that $\vvsym{\ell}_{10}$ contains a non-zero $\alpha$-component and is thus unphysical. Moreover, from our interpretation of $a_5$ and $a_6$ as arising from the physics of band bottoms, we expect a topological excitation in these two components to be much more energetically costly than those of the other matter fields. Hence we can also neglect $\vvsym{\ell}_{8}$ and $\vvsym{\ell}_{9}$. Thus only the top-left block of $K_r^{-1}$ is relevant for the statistics of low-lying physical quasiparticle excitations. Henceforth we shall restrict the meaning ``quasiparticle'' to those whose $\vvsym{\ell}$-vector is an integer combination of $\vvsym{\ell}_1$ through $\vvsym{\ell}_7$.

From $K_r^{-1}$ it can be seen that the system contains quasiparticles with non-trivial mutual statistics. In particular, \emph{there there are fermions having semionic mutual statistics} (i.e., a phase factor of $\pi$ when one quasiparticle winds around another), manifesting in, e.g., quasiparticles described by $\vvsym{\ell}_4$ and $\vvsym{\ell}_5$.

\begin{figure}
\subfigure[\label{fig:spinon}]{\includegraphics[scale=0.65]{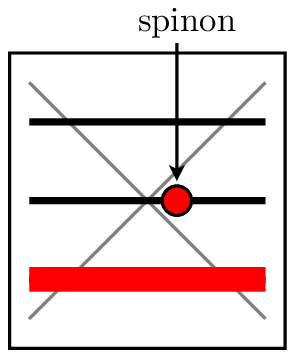}} \,
\subfigure[\label{fig:anti-spinon}]{\includegraphics[scale=0.65]{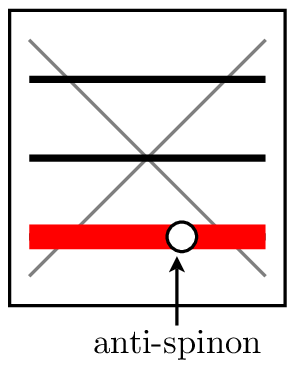}}\,
\subfigure[\label{fig:half-holon}]{\includegraphics[scale=0.65]{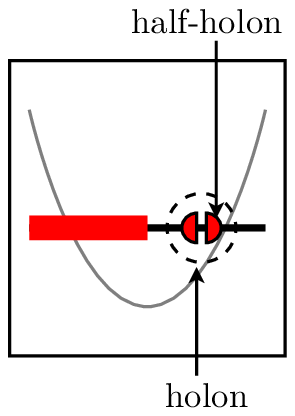}} \,
\subfigure[\label{fig:anti-half-holon}]{\includegraphics[scale=0.65]{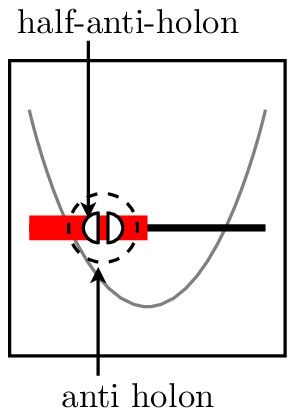}}
	\caption{\label{fig:QP} Physical interpretation of $\vvsym{\ell}$-vector: (a) a single ``$+1$'' in a spinon component identified as spinon; (b) a single ``$-1$'' in a spinon component identified  as anti-spinon; (c) a single ``$+1$'' (``$+2$'') in a holon component identified as half-holon (holon); and (d) a single ``$-1$'' (``$-2$'') in a holon component identified as anti-half-holon (anti-holon). The thick (red) horizontal lines indicate filled LLs that forms the ground state of the system, while the thin (black) horizontal lines indicate unfilled LLs.}
\end{figure}

The self-statistics and mutual statistics of different quasiparticles can be understood intuitively. Recall that our system is constructed by coupling integer and fractional quantum Hall states via a common constraint gauge field $\alpha$. If we assume that the different quantum Hall states are independent of each other, i.e., a ``charge'' in one matter-field component has trivial bosonic statistics with a ``charge'' in a different matter-field component, then \emph{the statistics of these quasiparticles can be read off by considering their underlying constituents}. For example, since $\vvsym{\ell}_{4}$ and $\vvsym{\ell}_{5}$ overlaps in one $\nu=1/2$ component, their mutual statistics is semionic. Similarly, since $\vvsym{\ell}_{4}$ overlaps with itself in two $\nu=1/2$ components, its self-statistics is fermionic.\footnote{For this intuitive picture to be accurate, the sign of the component must also be taken into account.} From this intuitive picture, it is evident that a ``$+1$'' in a spinon component in the $\vvsym{\ell}$-vector should be identified with a spinon excitation on top of the integer quantum Hall state that formed near the corresponding Dirac node, while a ``$+1$'' in a holon component in the $\vvsym{\ell}$-vector should be identified with \emph{half-holon} excitation on top of the $\nu=1/2$ quantum Hall state that formed near the corresponding band bottom. Similarly, a ``$-1$'' in a spinon (holon) component in the $\vvsym{\ell}$-vector should be identified as an anti-spinon (anti-half-holon). See Fig.~\ref{fig:QP} for illustration.

To discuss these quasiparticles further, it is useful to divide them into three classes. The first class consists of quasiparticles with spinon components only and will be referred to as ``spinon quasiparticles'' (SQP). The second class consists of quasiparticles with holon components only and will be referred to as ``holon quasiparticles'' (HQP). The remaining class consists of quasiparticles that have both spinon and holon components, and will be referred to as ``mixed quasiparticles'' (MQP). The first two classes can be constructed by compounding ``elementary'' quasiparticles of the same type. For SQP, the ``elementary'' quasiparticles are described by $\vvsym{\ell}$-vectors having exactly one ``$+1$'' component and one ``$-1$'' component in the spinon sector (e.g., the $\vvsym{\ell}_1, \vvsym{\ell}_2$, and $\vvsym{\ell}_3$ in Eq.~\ref{eq:lbasis}). For HQP, the ``elementary'' quasiparticles are described by $\vvsym{\ell}$-vectors having exactly one ``$+1$'' component and one ``$-1$'' component in the holon sector (e.g., the $\vvsym{\ell}_4, \vvsym{\ell}_5$, and $\vvsym{\ell}_6$ in Eq.~\ref{eq:lbasis}). As for the MQP, one can start with ``minimal'' quasiparticles with exactly one ``$+1$'' component in the spinon sector and one ``$+2$'' components in the holon sector, and build all MQP by compounding at least one such ``minimal'' quasiparticles together with zero or more ``elementary'' SQP and HQP. Alternatively, one may start with a second type of ``minimal'' quasiparticle in the MQP sector, which has exactly one ``$+1$'' component in the spinon sector and two ``$+1$'' components in the holon sector, and build all MQP by compounding at least one such ``minimal'' quasiparticles together with zero or more ``elementary'' SQP and HQP (note that the second-type of ``minimal'' MQP is simply a ``minimal'' MQP of the first type compounded with an ``elementary'' HQP. The introduction of two different types of ``minimal'' MQP will be clear in the following).

\begin{figure}
\begin{center}
\subfigure[\label{fig:SQP_elem}]{\includegraphics[scale=0.5]{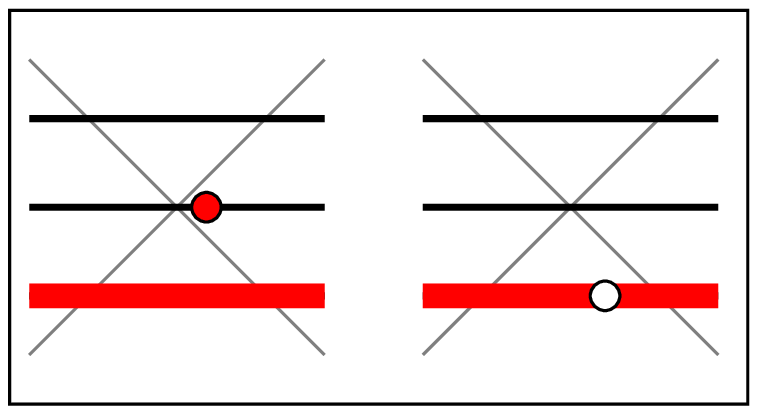}} \quad
\subfigure[\label{fig:HQP_elem}]{\includegraphics[scale=0.5]{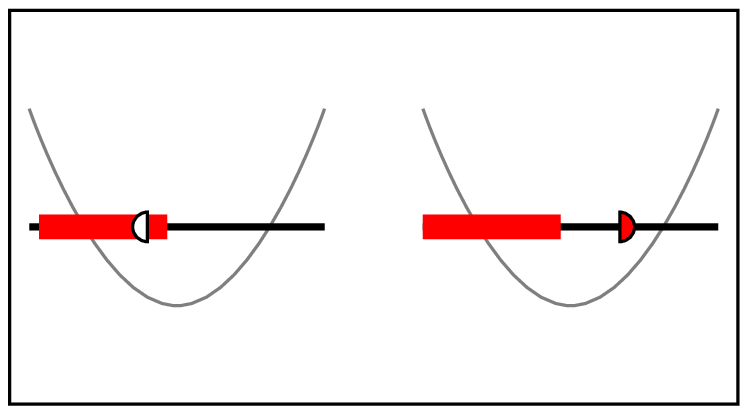}} \\
\subfigure[\label{fig:MQP_min_I}]{\includegraphics[scale=0.45]{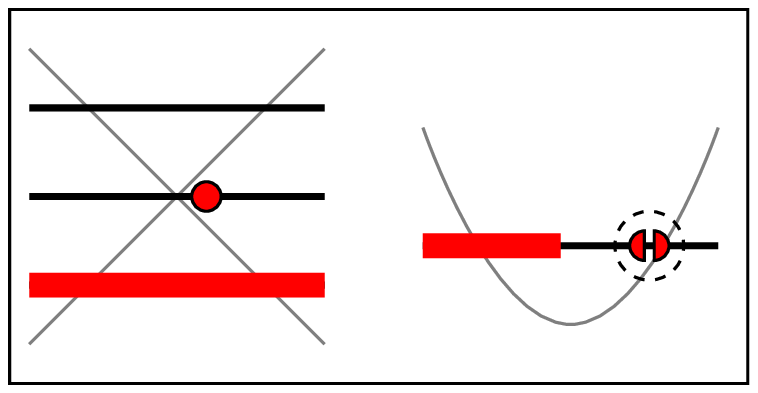}} \quad
\subfigure[\label{fig:MQP_min_II}]{\includegraphics[scale=0.45]{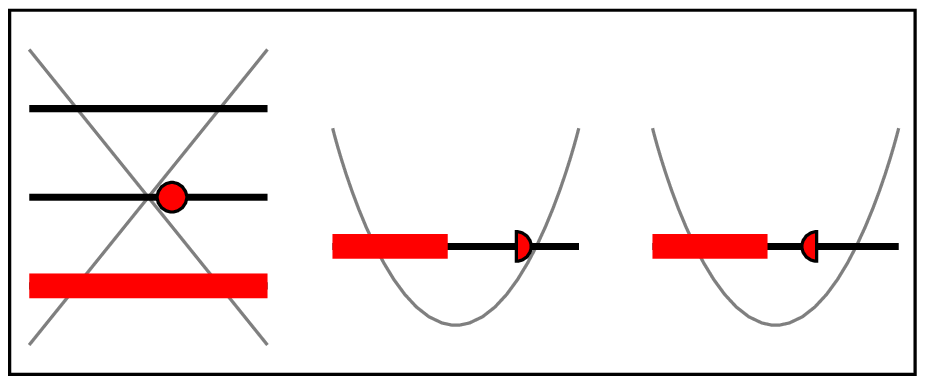}}
\caption{\label{fig:neutral_QP}Visualization of the (a) ``elementary'' SQP; (b) ``elementary'' HQP; (c) ``minimal'' MQP of the first type; and (d) ``minimal'' MQP of the second type.}
\end{center}
\end{figure}

These ``elementary'' and ``minimal'' quasiparticle excitations can be visualized in the following way: The ``elementary'' SQP can be visualized as a particle-hole excitation in the spinon quantum Hall levels, in which a spinon is removed from one Dirac node and added in another. The elementary HQP can be visualized as a particle-hole excitation in the holon quantum Hall levels, in which a \emph{half} holon is transferred from one band bottom to another. The minimal SQP can be visualized as adding both spinon and (half) holons into the original system. See Fig.~\ref{fig:neutral_QP} for illustrations.

\begin{table}
\caption{\label{tbl:qp} Self- and mutual- statistics of the ``elementary'' or ``minimal'' quasiparticles in the doped kagom\'e system. The adjective ``elementary'' or ``minimal'' are omitted but assumed in the table entries. The subscript I and II indicates the type of ``minimal'' MQP considered (see the main text for their definitions). When an entry contain multiple cases, both cases are possible but are realized by different quasiparticles in the respective sectors.}
\begin{ruledtabular}
\begin{tabular}{cccccc}
& & \multicolumn{4}{c}{Mutual Statistical Phase\footnotemark[2]} \\
\raisebox{1.5ex}[0cm][0cm]{Type}
	& \raisebox{1.5ex}[0cm][0cm]{Self-Statistics\footnotemark[1]}
	& SPH & HPH & MQP$_{\textrm{I}}$ & MQP$_{\textrm{II}}$\\
\hline
SQP & b & $2\pi$ & $2\pi$ & $2\pi$ & $2\pi$\\
HQP & f & $2\pi$ & $\pi$ or $2\pi$ & $2\pi$ & $\pi$ or $2\pi$\\
MQP$_{\textrm{I}}$ & f & $2\pi$ & $2\pi$ & $2\pi$ & $2\pi$\\
MQP$_{\textrm{II}}$ & b & $2\pi$ & $\pi$ or $2\pi$ & $2\pi$ & $\pi$ or $2\pi$
\end{tabular}
\end{ruledtabular}
\footnotetext[1]{b=bosonic, f=fermionic, s=semionic}
\footnotetext[2]{Phase angle accumulated when one quasiparticle winds around another, modulo $2\pi$.}
\end{table}

With this classification, the information on the self- and mutual- statistics of the quasiparticles contained in $K_r^{-1}$ can be summarized more transparently in terms of the self- and mutual- statistics of the ``elementary'' SQP, ``elementary'' HQP, and ``minimal'' MQP. The result is presented in Table~\ref{tbl:qp}.

\section{Quasiparticles---Quantum Numbers} \label{sect:qp_Qnum}

Since the quasiparticles have finite energy gaps and short-ranged interactions, they may carry well-defined quantum numbers. In particular, it is sensible to consider the $\vv{k}$ quantum numbers for these quasiparticles, since they arise from LLs that form near Dirac points or band bottoms with well-defined crystal momentum $\vv{k}$. Similarly, it is sensible to consider the $S_z$ quantum numbers for quasiparticles with spinon components. We shall see that this program can be carried out for ``elementary'' spinon quasiparticles and for the ``minimal'' mixed quasiparticles of first type, but not easily for the ``elementary'' holon quasiparticles and the ``minimal'' mixed quasiparticles of the second type.

Recall that we constructed a tight-binding model with doubled unit cell for the DSL ansatz. The unit cell is necessarily doubled because the DSL ansatz enclose a flux of $\pi$ within the original unit cell spanned by $\vv{r}_1/2 = \uv{x}$ and $\vv{r}_2 = (1/2) \uv{x} + (\sqrt{3}/2) \uv{y}$ (c.f.\@ Fig.~\ref{fig:R_lattice}), and hence the operators that corresponds to translation by $\uv{x}$, $T_x$, and the operator that corresponds to translation by $\vv{r}_2$, $T_{r2}$, do not commute in general (i.e., $[T_x,T_{r2}]\neq 0$), even though both commute with the mean-field tight-binding Hamiltonian. Consequently, single-spinon and single-holon states in the DSL ansatz generally form multi-dimensional irreducible representations under the joint action of $T_x$ and $T_{r2}$ (i.e., $T_x$ and $T_{r2}$ manifest as multi-dimensional matrices that cannot be simultaneously diagonalized when acting on these states), and cannot be labeled simply by a pair of numbers $(c_1,c_2)$ as in the ordinary case.\footnote{In the ordinary case, $(c_1,c_2)$ are simply eigenvalues of $T_x$ and $T_{r2}$, respectively, and are related to the crystal momentum $\vv{k}$ in the \emph{original} Brillouin zone via $\exp(i\vv{k}\cdot\uv{x}) = c_1$ and $\exp(i\vv{k}\cdot\vv{r}_2) = c_2$.} Furthermore, the matrices for $T_x$ and $T_{r2}$ will in general be \emph{$\alpha$-gauge-dependent}. However, when an even number of spinon and holon excitations are considered as a whole, the total phase accumulated when the particles circle around the original unit cell becomes a multiple of $2\pi$, and thus $[T_x,T_{r2}] =  0$ \emph{in such subspace}. Hence it is possible to reconstruct the crystal momentum in the original Brillouin zone if our attention is restricted to such states. The tool for reconstructing the crystal momentum in the original Brillouin zone is known as the \emph{projective symmetry group} (PSG).\cite{PSG:Wen} Physically, the gauge dependence of single-spinon and single-holon states indicate that they cannot be created \emph{alone}.

It can be checked that all SQP are composed of an even number of spinons and anti-spinons. The above discussion then implies that they carry well-defined $\vv{k}$ quantum numbers in the original Brillouin zone. To derive the transformational properties under $T_{x}$ and $T_{r2}$, we compute the transformation properties of the original spinon matter fields. The procedures for doing so have been described in details in Ref.~\onlinecite{DSL:Hermele}, here we shall just state the results.

\begin{figure}[htb]
\subfigure[\label{fig:spinon_label}]{\includegraphics[scale=0.9]{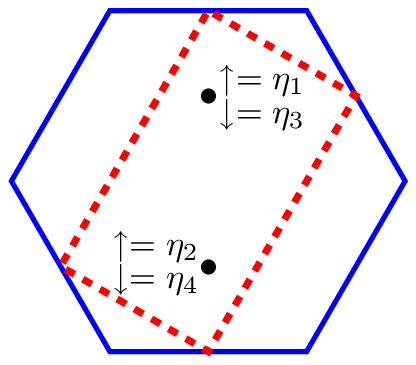}} \,
\subfigure[\label{fig:SQP_missing}]{\includegraphics[scale=0.7]{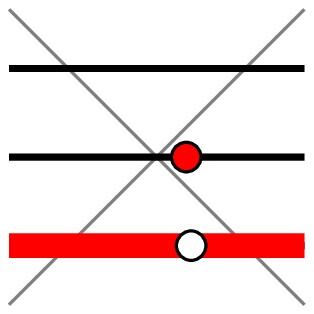}}\\
\subfigure[\label{fig:SQP_Sz}]{\includegraphics[scale=0.8]{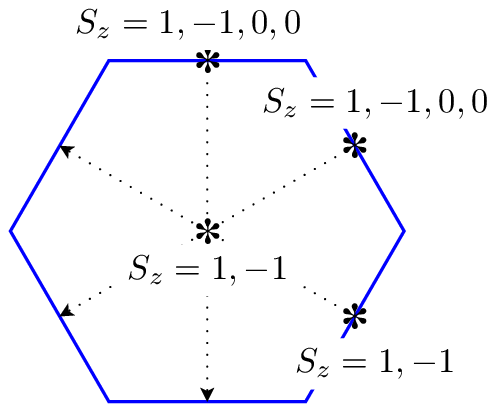}} \,
\subfigure[\label{fig:SQP_SU2}]{\includegraphics[scale=0.8]{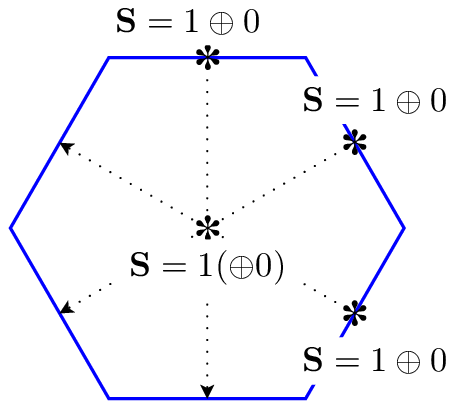}}
	\caption{\label{fig:SQP} (a) The labels for the four spinon topological excitations. (b) Physical interpretation of the ``missing states'' in the fixed $S_z$ quantization and doubled unit cell Chern--Simons formulation. (c) Spectrum of ``elementary'' SQP, with $\vv{k}$ and $S_z$ quantum number indicated, before restoring full symmetry. (d) Spectrum of ``elementary'' SQP after restoring the SU(2) symmetry by adding extra quasiparticles. The dotted arrows indicate equivalent $\vv{k}$-point upon translation by the \emph{original} reciprocal lattice vectors (spanned by $2 \vv{k}_1$ and $\vv{k}_2$ in Fig.~\ref{fig:k_lattice}). For dimension of the Brillouin zone, c.f.\@ Fig.~\ref{fig:k_lattice}.}
\end{figure}

Let $\eta_1, \ldots, \eta_4$ denote the topological excitations near the four (two $\vv{k}$-vectors and two spins) Dirac nodes as indicated in Fig.~\ref{fig:spinon_label}. Then, assuming that they have the same transformational properties as the underlying spinon fields at the same Dirac nodes,
\begin{equation} \label{eq:spinon_trans}
\begin{aligned}
T_x[\eta_1] & = e^{i\pi/12} \eta_2 \, , & T_{r2}[\eta_1] & = e^{i\pi/2} \eta_1 \, , \\
T_x[\eta_2] & = e^{11i\pi/12} \eta_1 \, , & T_{r2}[\eta_2] & = e^{-i\pi/2} \eta_2 \, , \\
T_x[\eta_3] & = e^{i\pi/12} \eta_4 \, , & T_{r2}[\eta_3] & = e^{i\pi/2} \eta_3 \, , \\
T_x[\eta_4] & = e^{11i\pi/12} \eta_3 \, , & T_{r2}[\eta_4] & = e^{-i\pi/2} \eta_4 \, .
\end{aligned}
\end{equation}
Furthermore, we assume that $T_x$ and $T_{r2}$ satisfy the generic conjugation and composition laws:
\begin{equation}
\begin{aligned}
T[\psi^*] & = (T[\psi])^* \, , & T[\psi\cdot \psi'] & = T[\psi]\cdot T[\psi'] \, . 
\end{aligned}
\end{equation}
where $\psi$, $\psi'$ denotes generic quasiparticle states, $\psi^*$ denotes an anti-particle of $\psi$, and $\psi \cdot \psi'$ denotes a bound state composed of $\psi$ and $\psi'$.

A general basis for ``elementary'' SQP is spanned by $\eta_i \eta_j^*$ with $i \neq j$. There are twelve distinct ``elementary'' SQP, which form six \emph{reducible} representations under $T_x$ and $T_{r2}$. Upon diagonalization, the resulting ``elementary'' SQP in the new basis each carry distinct $S_z$ and $\vv{k}$ (in the original Brillouin zone) quantum numbers. These are summarized in Fig.~\ref{fig:SQP_Sz}.

Notice that Fig.~\ref{fig:SQP_Sz} is somewhat unsettling. First, even though we have not performed a PSG study on rotation operators, intuition on rotation symmetry suggests that there should be four states (with $S_z = 1,-1,0, \textrm{and } 0$) located at $\vv{k} = (\pi,-\pi/\sqrt{3})$. Second, although our Chern--Simons theory is formulated with a fixed quantization axis for spin, the $SU(2)$ spin-rotation symmetry should remain unbroken. Therefore, the $S_z$ eigenvalues should organize into $SU(2)$ representations for each $\vv{k}$ value. While this is true for $\vv{k} = (\pi,\pi/\sqrt{3})$ and $\vv{k} = (0,2\pi/\sqrt{3})$, where the ``elementary'' SQP form $1 \oplus 0$ representations, the same does not hold for $\vv{k} = (\pi,-\pi/\sqrt{3})$ and $\vv{k} = (0,0)$.

The two issues mentioned above indicate that some topological excitations are lost in our formulation. In other words, \emph{there are topological excitations that have trivial $S_z$ quantum numbers but non-trivial $\vv{S}$ quantum numbers. Similarly, there are topological excitations that have trivial $\vv{k}$ quantum numbers in the reduced Brillouin zone but non-trivial $\vv{k}$ quantum numbers in the original Brillouin zone.}  Physically, the original of these missing excitations can be understood as follows: in the hydrodynamic approach, an $\vvsym{\ell}$-vector with a single ``$+1$'' in a spinon component represent a spinon at a Dirac node, while $\vvsym{\ell}$-vector with a single ``$-1$'' in a spinon component represent an anti-spinon at a Dirac node. The previously defined set of $\vvsym{\ell}$-vectors that characterized the ``elementary'' SQP  fail to captured an excitonic state in which a spinon is excited from a filled LL to an empty LL, thus leaving an anti-spinon behind (see Fig.~\ref{fig:SQP_missing} for an illustration), which precisely carry trivial $S_z$ quantum numbers and transoform trivially under $T_{r1}$ and $T_{r2}$. Note that there are four possible excitonic states of this form, hence we expect four states to be added. In our Chern--Simons formulation, these excitations may be disguised as combinations of density operators ($\sim \partial \vv{c}$).

From Fig.~\ref{fig:SQP_Sz} and the forgoing discussions, it is evident that extra states should be added at $\vv{k} = (\pi,\pi/\sqrt{3})$ and $\vv{k} = (0,0)$, so that the states at $\vv{k} = (0,0)$ and $\vv{k} = (\pi,\pi/\sqrt{3})$ each form a $1 \oplus 0$ representation of $SU(2)$. The final result after making this reparation is shown in Fig.~\ref{fig:SQP}(c). Formally, the same result can be reached if we allow objects of the form $\eta_i \eta_i^*$ to be counted as elementary SQP, then apply Eq.~\ref{eq:spinon_trans} and the procedure of diagonalization as before in this extended basis.

Observe that the ``elementary'' spinon SQP (and hence the entire SQP sector) all carry integer spins. However, we also know that a conventional superconductor contains spin-1/2 fermionic excitations (i.e., the Bogoliubov quasiparticles). From our assignment of $S_z$ quantum number and from the table of quasiparticle statistics Table~\ref{tbl:qp}, it is evident that the ``minimal'' MQP of the first type play the role the these Bogoliubov quasiparticles in the doped kagom\'e system. In contrast, minimal MQP of the second type are \emph{spin-1/2 quasiparticles that carry bosonic statistics} and hence is another distinctive signatures of this exotic superconductor.

Since a ``minimal'' MQP of the first type can be treated as a bound state of a spinon and a holon (c.f.\@ Fig.~\ref{fig:MQP_min_I}), the $\vv{k}$ quantum number in the original Brillouin zone are again well-defined for them. To construct their quantum numbers, we need to know how holons transform under $T_{x}$ and $T_{r2}$. Let $\varphi_1, \ldots, \varphi_4$ denotes the \emph{half-holon} excitations near the four holon band bottom as indicated in Fig.~\ref{fig:spinon_label}, such that $\varphi_1^2, \ldots, \varphi_4^2$ denotes the corresponding \emph{holon} excitations (c.f.\@ Fig.~\ref{fig:half-holon}). Following the same procedure that produces Eq.~\ref{eq:spinon_trans}, we obtain the transformation laws:

\begin{equation} \label{eq:holon_trans}
\begin{aligned}
T_x[\varphi_1^2] & = \varphi_4^2 \, , 
	& T_{r2}[\varphi_1^2] & = e^{i\pi/6} \varphi_1^2 \, , \\
T_x[\varphi_2^2] & = \varphi_3^2 \, , 
	& T_{r2}[\varphi_2^2] & = e^{-i\pi/6} \varphi_2^2 \, , \\
T_x[\varphi_3^2] & = e^{-i\pi/3} \varphi_2^2 \, , 
	& T_{r2}[\varphi_3^2] & = e^{5i\pi/6} \varphi_3^2 \, , \\
T_x[\varphi_4^2] & = e^{i\pi/3} \varphi_1^2 \, , 
	& T_{r2}[\varphi_4^2] & = e^{-5i\pi/6} \varphi_4^2 \, .
\end{aligned}
\end{equation}

\begin{figure}
\subfigure[\label{fig:holon_label}]{\includegraphics[scale=0.9]{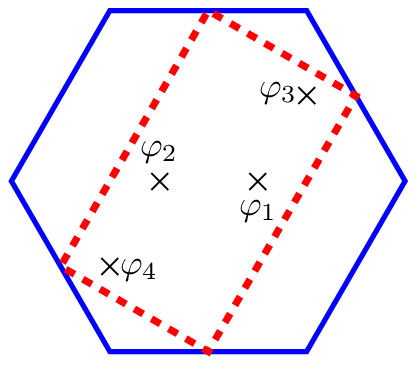}} \,
\subfigure[\label{fig:MQP_SU2}]{\includegraphics[scale=0.9]{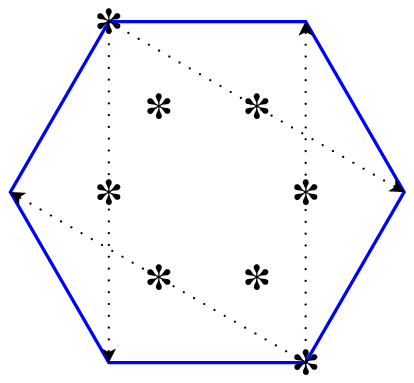}}
	\caption{\label{fig:MQP} (a) The labels for the four holon excitations. (b) Spectrum of ``elementary'' MQP, with $\vv{k}$ quantum number indicated. Each point in $\vv{k}$ space forms a $\vv{S} = 1/2$ representation in spin. The dotted arrows indicate equivalent $\vv{k}$-point upon translation by the \emph{original} reciprocal lattice vectors.}
\end{figure}

A general basis for ``minimal'' MQP of the first type is spanned by $\eta_i \varphi_j^2$. There are sixteen distinct first-type ``minimal'' MQP, which form eight \emph{reducible} representations under $T_x$ and $T_{r2}$. Upon diagonalization, the resulting first-type ``minimal'' MQP in the new basis each carry distinct $S_z$ and $\vv{k}$ (in the original Brillouin zone) quantum numbers, and the full $SU(2)$ representation in spin can be recovered trivially by combining spin-up and spin-down states. The final results are summarized in Fig.~\ref{fig:MQP_SU2}.

Having considered the SQP sector and the ``minimal'' MQP of the first type, one may attempt to carry out similar analysis for the HQP sector and for the ``minimal'' MQP of the second type. However, in doing so, issues arise from the fractionalization of holons into half-holons. Recall that in deriving the transformational rules of the quasiparticles, we identify the components of $\vvsym{\ell}$ as being spinon and holon excitations, and assume that these excitations carry the same quantum numbers as the underlying spinons and holons that form the LLs in the first place. However, the HQP sector and the ``minimal'' MQP of the second type are bound states that involve \emph{half-holons}, whose quantum numbers cannot be directly inferred from the underlying spinons and holons. More concretely, we need to know the transformation laws $T[\varphi_i \varphi_j^*]$ for half-holon--anti-half-holon pairs $\varphi_i \varphi_j^*$ in order to construct their quantum numbers, but we only have information about transformation laws $T[\varphi_j^2]$ of holon excitation $\varphi_j^2$. 

It is far from clear how $T[\varphi_i \varphi_j^*]$ can be related to $T[\varphi_j^2]$. The answer for such question may even be non-unique. We have already seen an analogous situation in the forgoing discussion: while the spinon--anti-spinon pairs $\eta_i\eta_j^*$ have well-defined gauge-invariant $\vv{k}$ quantum numbers in the original Brillouin zone, the single spinons $\eta_i$ form gauge-dependent two-dimensional representations under $T_x$.

The possible ambiguity in the transformation law $T[\varphi_i \varphi_j^*]$ of half-holon--anti-half-holon pairs $\varphi_i \varphi_j^*$ signifies that it may not be possible to produce these quasiparticles \emph{alone}. Although a half-holon--anti-half-holon pair can be thought of as resulted from removing a half-holon from one band bottom and adding one in another, it is not clear that the process can be done in via single half-holon tunneling. This is analogous to the case when two fractional quantum Hall system are separated by a constriction, where it is only possible to tunnel physical electrons.\cite{QHE:Wen}

Combining the results from Sect.~\ref{sect:qp_stat} and \ref{sect:qp_Qnum}, we see that there are two very different class of quasiparticle excitations in the doped kagome system---which can be termed as ``conventional'' and ``exotic,'' respectively. The ``conventional'' class consists of quasiparticles that can be created alone, which carry well-defined crystal momentum $\vv{k}$ in the original Brillouin zone and possess conventional (fermionic or bosonic) statistics. These include the spinon particle-holes, the holon (but not half-holon) particle-holes, the ``minimal'' mixed quasiparticles of the first type (a.k.a.\@ the ``Bogoliubov quasiparticles''), and their composites. 
In contrast, the ``exotic'' class consists of quasiparticle that cannot be created alone, whose crystal momentum may not be well-defined, and whose statistics may be fractional. These include the half-holon particle-holes and the ``minimal'' mixed quasiparticles of the second type (which are ``Bogoliubov quasiparticles'' dressed with a half-holon particle-hole). In terms of the \emph{underlying electronic system}, the former class are excitations that are \emph{local} in terms of the \emph{underlying} electron operators $\op{c}$ and $\opdag{c}$, while the latter class are excitations that are \emph{non-local} in terms of  $\op{c}$ and $\opdag{c}$.

It should be warned that questions regarding the energetics (and hence stability) of the quasiparticles have not been touched in Sect.~\ref{sect:qp_stat} and \ref{sect:qp_Qnum}. In particular, it is not clear whether the bosonic or the fermionic spin-1/2 excitation has a lower energy. Though this information is in principle contained in the Maxwell term Eq.~\ref{eq:QHE_Maxwell}, to obtain it requires a detailed consideration of the short-distance physics in the $t$--$J$ model, and is beyond the scope of this paper. 

\section{An alternative derivation by eliminating the auxiliary field} \label{sect:elif}

It is a curious result that in Eq.~\ref{eq:invKr}, once the unphysical $\vvsym{\ell}_{10}$ is removed from the spectrum, the quasiparticle represented by $\vvsym{\ell}_{9}$ becomes purely bosonic (i.e., having trivial bosonic mutual statistics with all other quasiparticles and trivial bosonic self-statistics). This suggests that $\vvsym{\ell}_{9}$ corresponds to some local density excitation of the system and thus should not be regarded as topological. Moreover, the procedure of first treating $\vvsym{\ell}_{10}$ as part of the spectrum in computing $K_r^{-1}$ and then removing this degree of freedom at the very end of the calculation seems somewhat dubious. Recall that the gauge field $\alpha^\mu$ is introduced to enforce the occupation constraint Eq.~\ref{eq:constraint}. This gauge field is thus an \emph{auxiliary} field that is void of self-dynamics (i.e., the term $\eadb{\alpha}{\alpha}$ vanishes) and topologically trivial (i.e., the $\alpha$-component of $\vvsym{\ell}$ must be zero). Therefore, one may attempt to \emph{re-derive the previous results by eliminating this $\alpha$ field right at the beginning by enforcing the constraint directly}. This can indeed be done, as we shall show in the following.

Recall that the EOM with respect to $\alpha^\mu$ leads to the constraint equation Eq.~\ref{eq:alpha_EOM} in the Chern--Simons formulation. From this, one may argue that \emph{the effect of introducing the $\alpha$ field can alternatively be produced by setting $\sum_I a_I^\mu + \sum_J b_J^\mu = 0$ directly}. To do so, we perform a two-step transformation on the Lagrangian Eq.~\ref{eq:Lg}. First, we set:
\begin{align} \label{eq:elim_trans1}
{a'_6}^\mu & = \sum_I a_I^\mu &\textrm{and}& 
	& {a'_I}^\mu = a_I^\mu \ \textrm{for $I \neq 6$} \, ;\\
{b'_1}^\mu & = \sum_J a_J^\mu &\textrm{and}& 
	& {b'_J}^\mu = b_J^\mu \ \textrm{for $J \neq 1$} \, .
\end{align}
Then the constraint becomes ${a'_6}^\mu + {b'_1}^\mu = 0$, which we enforce directly by setting:
\begin{equation} \label{eq:elim_trans2}
\rho^\mu = -{a'_6}^\mu = {b'_1}^\mu \, ,
\end{equation}
thus eliminating one variable.

Note that since the $\alpha$ field appears in Eq.~\ref{eq:Lg} only through the term $\eadb{(\sum_I a_I + \sum_J b_J)}{\alpha}$, it got dropped out of the transformed Chern--Simons Lagrangian. Letting $\elif{\vv{c}} = (\rho; a'_1, \ldots, a'_5; b'_2, \ldots, b'_4)$, which is a column vector of only nine (as opposed to eleven) gauge fields, Eq.~\ref{eq:Lg} becomes:
\begin{equation} \label{eq:elim_Lg}
\Lg = \frac{1}{4\pi} \eadb{\elif{\vv{c}}^T \elif{K}}{\elif{\vv{c}}} 
	- \frac{e}{2\pi} \eadb{(\elif{\vv{q}} \cdot \elif{\vv{c}})}{A}
	+ (\elif{\vvsym{\ell}} \cdot \elif{\vv{c}})_\mu j_V^\mu + \ldots \, ,
\end{equation}
where $\elif{\vv{q}} = (1;\Zeros_{1,5};\Zeros_{1,3})^T$ is the transformed charge vector, and $\elif{K}$ is the transformed $K$-matrix:
\begin{equation} \label{eq:elim_K}
\elif{K} \!=\! \left( \begin{array}{c|ccccc|ccc}
3		& 1 & 1 & 1 & 1 & 1 		& -2 & -2 & -2 \\ \hline
1		& 0 & 1 & 1 & 1 & 1 		& & & \\
1 	& 1 & 0 & 1 & 1 & 1 		&	& & \\
1		& 1 & 1 & 0 & 1 & 1 		&	\multicolumn{3}{c}{\Zeros_{5,3}}\\
1 	& 1 & 1 & 1 & 0 & 1 		&	& & \\
1 	& 1 & 1 & 1 & 1 & 2 		& & & \\ \hline
-2	& & & & & 							&	4 & 2 & 2 \\
-2	& & \multicolumn{3}{c}{\Zeros_{3,5}} & & 2 & 4 & 2 \\
-2	& & & & & 							& 2 & 2 & 4
\end{array} \right)
\end{equation}

As for topological excitations, from the transformation between $\vv{c}$ and $\elif{\vv{c}}$, it can be seen that the correspondence between $\vvsym{\ell}$ and $\elif{\vvsym{\ell}}$ reads:
\begin{equation} \label{eq:lcorr} \begin{gathered}
\vvsym{\ell} = (0; n_{a1}, \ldots n_{a6}; n_{b1}, \ldots n_{b4})^T \\
\Updownarrow \\
\begin{aligned}
\elif{\vvsym{\ell}} & =(n_{b1}\!-\!n_{a6}; n_{a1}\!-\!n_{a6},\ldots,n_{a5}\!-\!n_{a6};\\
 & \quad n_{b2}\!-\!n_{b1},\ldots, n_{b4}\!-\!n_{b1})^T \end{aligned}
\end{gathered}\end{equation}
Hence, $\vvsym{\ell}$ is an integer vector if and only if $\elif{\vvsym{\ell}}$ is also an integer vector. Moreover, from Eq.~\ref{eq:lcorr} it can be seen that $\vvsym{\ell}_{9}$ is mapped to $\elif{\vvsym{\ell}} = \vv{0}$,\footnote{More generally, given $\elif{\vvsym{\ell}}$ in the transformed basis, the corresponding $\vvsym{\ell}$ is determined up to multiples of $\vvsym{\ell}_{9}$.} which is consistent with our previous argument that the quasiparticle corresponding to $\vvsym{\ell}_{9}$ is purely bosonic and hence should be considered as non-topological.

Although $K$ and $\elif{K}$ look rather different superficially,\footnote{It can even be checked that $K$ contains irrational eigenvalues that are \emph{not} eigenvalues of $\elif{K}$.} all the major conclusions from Sect.~\ref{sect:zero-mode}--\ref{sect:qp_Qnum} can be reproduced with $\elif{K}$. In particular, we shall check that the existence of a single gapless mode, the $hc/4e$ flux through a minimal vortex, and the semionic quasiparticle statistics can all be obtained from $\elif{K}$.

It is easy to check that $\elif{K}$ has exactly one zero eigenvalue, with $\elif{\vv{p}}_0 = (4;-2\Ones_{1,4},2;\Ones_{1,3})^T$ its eigenvector. Using the transformation equations Eq.~\ref{eq:elim_trans1}--\ref{eq:elim_trans2}, we see that this eigenvector corresponds precisely to the eigenvector $\vv{p}_0$ we found in Sect.~\ref{sect:zero-mode}. Thus, again we conclude that the system contains a gapless mode associated with superconductivity, and that this gapless mode can be interpreted as fluctuations of all spinons and holons species whose ratio is matched (through their common coupling to the gauge field $\alpha^\mu$).

Moreover, the amount of magnetic flux that passes through a physical vortex is still described by Eq.~\ref{eq:Bflux} upon the obvious modifications. Since $\elif{\vv{q}} \cdot \elif{\vv{p}_0} = 4$, we recover the conclusion that a minimal physical vortex carries a flux of $hc/4e$. Furthermore, it can be checked that $\vvsym{\ell}\cdot\vv{p}_0 = \elif{\vvsym{\ell}}\cdot\elif{\vv{p}}_0$ for $\vvsym{\ell}$, $\elif{\vvsym{\ell}}$ satisfying the correspondence Eq.~\ref{eq:lcorr}. Hence the flux carried by a vortex calculated from $\elif{K}$ agrees with the value calculated from $K$.

As before, the quasiparticle excitations (which are EM-neutral, short-ranged interacting, and have finite energy gaps) are characterized by the condition that $\elif{\vvsym{\ell}} \cdot \elif{\vv{p}}_0 = 0$, which defines an eight-dimensional subspace of the nine-dimensional space in this case. The $K$-matrix restricted to this subspace, $\elif{K}_r$, is invertible. We may choose a basis for this subspace that corresponds to the basis choice Eq.~\ref{eq:lbasis} in the original representation. Explicitly,
\begin{equation} \label{eq:elim_lbasis}
\begin{aligned}
\elif{\vvsym{\ell}}_{1} & = (0; 0,-1,1,0,0;\Zeros_{1,3})^T \\
\elif{\vvsym{\ell}}_{2} & = (0; 0,-1,0,1,0;\Zeros_{1,3})^T \\
\elif{\vvsym{\ell}}_{3} & = (0; 0,-1,0,0,1;\Zeros_{1,3})^T \\
\elif{\vvsym{\ell}}_{4} & = (0; 0,\Zeros_{1,4};0,1,-1)^T \\
\elif{\vvsym{\ell}}_{5} & = (0; 0,\Zeros_{1,4};1,0,-1)^T \\
\elif{\vvsym{\ell}}_{6} & = (1; 0,\Zeros_{1,4};-1,-1,-2)^T \\
\elif{\vvsym{\ell}}_{7} & = (0; 0,0,1,0,0; 1,1,0)^T \\
\elif{\vvsym{\ell}}_{8} & = (0; 1,1,0,0,0;\Zeros_{1,3})^T 
\end{aligned}
\end{equation}

Then, it can be checked that:
\begin{equation} \label{eq:elim_invKr}
\elif{K}_r^{-1} \!=\! \left( \begin{array}{cccccccc}
-2 & -1 & -1 			& & & 							& -1 & 1 \\
-1 & -2 & -1 			& & \Zeros_{3,3} & 	& 0 & 1 \\
-1 & -1 & -2 			& & & 							& 0 & 1 \\
& & 							& 1 & 1/2 & 1/2 		& 1/2 & 0 \\
& \Zeros_{3,3} & 	& 1/2 & 1 & 1/2 		& 1/2 & 0 \\
& & 							& 1/2 & 1/2 & 1 		& 0 & 0 \\
-1 & 0 & 0 				& 1/2 & 1/2 & 0 		& 0 & 0 \\
1 & 1 & 1 				& 0 & 0 & 0 				& 0 & 0
\end{array} \right)
\end{equation}
in agreement with the results in Sect.~\ref{sect:qp_stat}.

\section{Conclusions} \label{sect:conclude}

In this paper we have considered the theory of a doped spin-1/2 kagom\'e lattice described by the $t$--$J$ model. We start with the slave-boson theory and the assumption that the undoped system is described by the $U(1)$ Dirac spin liquid, from which we argued that the doped system is analogous to a coupled quantum Hall system, with the role of the external magnetic field in the usual case taken up by an emergent gauge field $\alpha$. The analogy with quantum Hall systems compels us to introduce the Chern--Simons theory as an effective description of the low-energy physics of the system. This allows us to describe the superconductivity, the physical vortices, and the electromagnetically neutral quasiparticles in a unified mathematical framework. We show that there are two alternative Chern--Simons theories that produce identical results---one with the auxiliary field $\alpha$ kept until the end, and the other with the auxiliary field and a redundant dual matter field eliminated at the beginning.

In our scenario, the coupled quantum Hall system consists of four species of spinons and four species of holons at low energy. We show that such system exhibit superconductivity and that the flux carried by a minimal vortex is $hc/4e$. The system also contains fermionic quasiparticles with semionic mutual statistics, and bosonic spin-1/2 quasiparticle. As for the quantum numbers carried by the quasiparticles, we analyzed the spinon sector in details and found that it is possible to recover the full $SU(2)$ and (un-enlarged) lattice symmetry of the ``elementary'' quasiparticles in this sector, upon the inclusion of quasiparticles that are not easily represented in the original fixed-spin-quantization-axis, enlarged-unit-cell description. The same classification of quantum numbers are also carried out for the spin-1/2 fermionic quasiparticles, which are the analog of Bogoliubov quasiparticles in our exotic superconductor.

In this paper we have argued that the doped spin-1/2 kagom\'e system may exhibit exotic superconductivity that is higher unconventional. However, it should be remarked we have presented only one possible scenario for the doped kagom\'e system. For example, it is possible that the ground state of the undoped system is a valence bond solid\cite{VBS:Nikolic} and hence invalidate our analysis. Furthermore, experimentally realizing the idealized system considered considered in this paper may involve considerable difficulties. For instance, in the case of Herbertsmithite, it is known that the substitution between Cu and Zn atoms can be as big as ~5\%.\cite{Expt:Vries} It is our hope that this paper will generate further interests in the doped spin-1/2 kagom\'e system, as well as other systems that may exhibit anlogous exotic superconducting machanisms, both experimentally and theoretically.

\begin{acknowledgments}
We thank Ying Ran for discussions. This research is partially supported by NSF Grant No.\@ DMR-0804040 and DMR-0706078.
\end{acknowledgments}

\bibliographystyle{apsrev}
\bibliography{kagome_PRB}


\end{document}